# Text as data: a machine learning-based approach to measuring uncertainty

Rickard Nyman[1] and Paul Ormerod[2]

June 2020

***Abstract***   *The Economic Policy Uncertainty index had gained considerable traction with both academics and policy practitioners. Here, we analyse news feed data to construct a simple, general measure of uncertainty in the United States using a highly cited machine learning methodology. Over the period January 1996 through May 2020, we show that the series unequivocally Granger-causes the EPU and there is no Granger-causality in the reverse direction*

1.  **Introduction**

Following the seminal work of Baker et al. (2016) the Economic Policy Uncertainty index (EPU) has gained considerable influence. The original paper now has almost 4,000 citations according to Google Scholar. The index itself appears in speeches and papers by central bankers.

The website dedicated to the EPU (https://www.policyuncertainty.com/index.html) offers two different versions of the data. The first is based purely on a quantification of newspaper coverage of policy-related economic uncertainty. For the United States, this is based on a monthly search of 10 large newspapers. We refer to this below as EPUNEWS.

The second includes EPUNEWS as a component, but also incorporates information on tax code information and on disagreement amongst economic forecasters in the database published by the Federal Reserve Bank of Philadelphia's Survey of Professional Forecasters. We refer to this as EPUGEN (for "general"). Full details of the construction of both indices are provided on the website cited above.

The indices do not simply provide information about the current level of economic policy uncertainty. The website states that: "A significant dynamic relationship exists between our economic policy uncertainty index and real macroeconomic variables. We find that an increase in economic policy uncertainty as measured by our index foreshadows a decline in economic growth and employment in the following months".

---

[1] Centre for Decision Making Uncertainty, University College London and Algorithmic Economics Ltd
rickard.nyman.11@ucl.ac.uk)
[2] Department of Computer Science, University College London and Algorithmic Economics Ltd
p.ormerod@ucl.ac.uk

Here, we report a measure of uncertainty (UNCERT) which Granger-causes both EPUNEWS and EPUGEN in the United States over the period January 1996 through May 2020. In other words, changes in this measure systematically lead changes in both EPUNEWS and EPUGEN. There is no evidence of causation from the EPU indices to UNCERT.

UNCERT is constructed from the Reuters newsfeed for the United States using the unsupervised learning algorithm for obtaining vector representations for words developed by Pennington et al. (2014)[3].

Section 2 describes the construction of UNCERT and compares the three variables discussed above. Section 3 provides a summary of the results. An extended version of the results is set out in the Appendix.

## 2. The data series

Over the course of the most recent decade, important advances have been made in machine learning in the conversion of text into some form of quantitative representation.

A recent paper in the Journal of Economic Literature by Gentzkow et al. (2019) draws attention to the potential which these algorithms create for economists and other quantitative social scientists: "New technologies have made available vast quantities of digital text, recording an ever-increasing share of human interaction, communication, and culture. For social scientists, the information encoded in text is a rich complement to the more structured kinds of data traditionally used in research" (p.535).

We use an approach which has become standard in machine learning known as GloVe (Pennington et al. op.cit.). A clear overview, with a description of how to download and use the method, is given at https://nlp.stanford.edu/projects/glove/.

The authors assemble a very large corpus of words from various sources. We use the one described on the GloVe website as Common Crawl (glove.42B.300d.zip). A co-occurrence matrix is constructed, which describes how frequently pairs of words co-occur with each other in any given corpus.

The referenced webpage above states: "The training objective of GloVe is to learn word vectors such that their dot product equals the logarithm of the word's probability of co-occurrence. Owing to the fact that the logarithm of a ratio equals the difference of logarithms, this objective associates (the logarithm of) ratios of co-occurrence probabilities with vector differences in the word vector space".

---

[3] a paper which, incidentally, has over 13,000 citations

The eventual output of the process is that every word in the corpus has a unique n-dimensional vector associated with it. The elements of each vector are real valued numbers which essentially describe the closeness of the word to all other words in the corpus. This description is perforce rather imprecise. It is only intended to give a broad non-technical indication of what is going on. Full technical details are in Pennington et al. (op.cit.).

We apply the algorithm to the Reuters newsfeed over the period 1 January 1996 through 31 May 2020. To restrict the analysis to the US, we analyse all stories published by the New York and Washington offices, amounting to a total of 2,540,233 articles. The basic analysis is carried out on a daily basis, and the results aggregated onto a monthly frequency.

Our basic approach is simply to count the number of times the word "uncertainty" appears in the Reuters newsfeed each day. But we add to this word the four closest words to it, identified by the GloVe methodology. These are: "uncertainties", "uncertain", "unpredictability" and "ambiguity".

We then scale the raw data by counting the number of articles that mention at least one of the words, divided by the total number of articles. The scale is therefore the proportion of articles that matches the keyword search.

The closest word to "uncertainty" (except of course for the word itself) is "uncertainties". The Euclidean distance between the vector associated with "uncertainty" and the vector associated with "uncertainties" is 5.40. Of the 1.9 million words in the corpus, the Euclidean distance to the one furthest away is 17.70. The median is 8.64 and the standard deviation 0.68.

Figure 1 plots the Euclidean distance of the 200 words closest to "uncertainty".

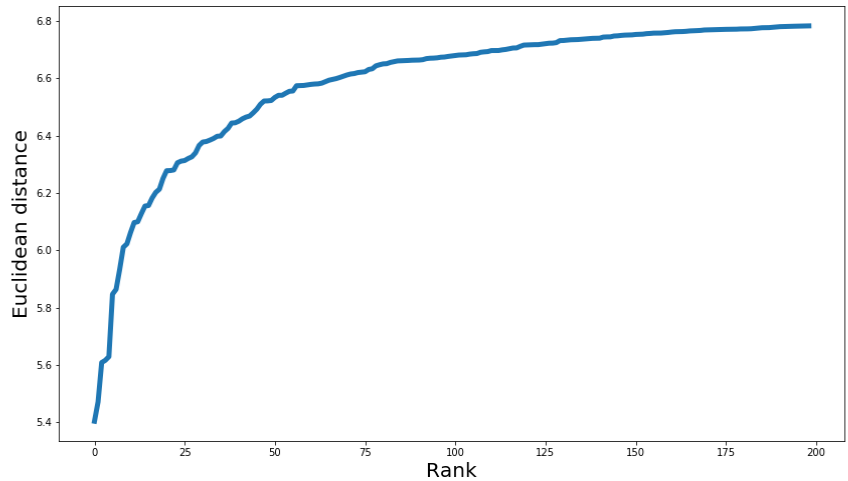

**Figure 1**  *Euclidean distance from the GloVe vector of "uncertainty" of the 200 words nearest to it*

As can be seen, the distance rises quite rapidly, and we select, as mentioned, the four nearest.

The monthly count of the five words is plotted with EUNEWS in Figure 2 for comparison, with both series put on the same units of measurement to enable this to be illustrated easily.

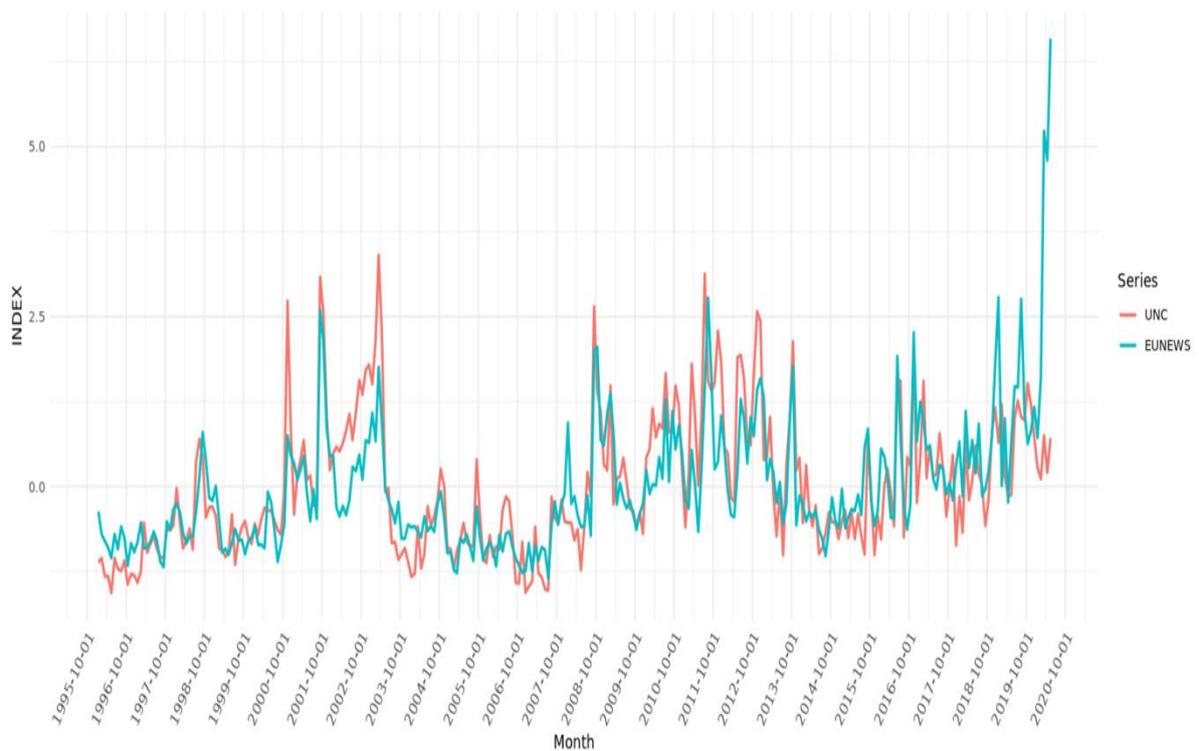

**Figure 2**  *The Uncertainty Index derived from the Reuters newsfeed and the Economic Policy Uncertainty Index based solely on news, January 1996 – May 2020*

In general, the two series move closely together. Uncertainty was high in the aftermath of the dot com boom in the early 2000s, and again during the financial crisis. After a temporary fall, it rose again in the early 2010s. The economic recovery in America, whilst stronger than it was in Europe, was weak by historical standards. Uncertainty appears to have been a factor in this. There is a marked divergence between the two series in April and May 2020, the period of the Covid crisis. One possible explanation is that although output dropped dramatically, the balance sheets of major companies were in general sufficiently strong to dampen doubts about any substantial defaults.

### 3. Results

We carry out Granger causality tests between UNCERT and, separately, both EUNEWS and EUGEN.

We do this both over the full sample period January 1996 through May 2020, and over each of two sub-samples, January 1996 through December 2007 and January 2008 through May 2020. The main reason is to check is there is any difference in the results in the pre- and post-financial crisis periods, though by coincidence this split gives sub-samples of very similar length.

We use the methodology described in Toda and Yamamoto (1995). In outline, in investigating Granger causality between any two series, this is as follows:

1. Check the order of integration of the two series using Augmented Dickey-Fuller and the Kwiatowski-Phillips-Schmidt-Shin tests. Let $m$ be the maximum order of integration found.

2. Specify the VAR model *using the data in levelled* form whatever was found in step 1 determine the number of lags to use with standard method. We use the Akaike Information Criteria

3. Check the stability of the VAR using OLS-CUSUM plots

4. Test for autocorrelation of residuals. If autocorrelation is found, increase the number of lags until it goes away. We use the multivariate Portmanteau- and Breusch-Godfrey tests for serially correlated errors. Let p be the number of lags then used

5. Add m extra lags of each variable to the VAR

6. Perform Wald tests with null being that the first p lags of the independent variable have coefficients equal to 0. If this is rejected, we have evidence of Granger-causality from the independent to dependent variable.

In the Appendix we describe the specific functions used in R and some further details of the results. A complete set of results, including the R commands used at each stage of the process is available from Nyman (rickard.nyman.11@ucl.ac.uk)

**Table 1**      **P-values in tests of Granger causality between UNCERT and EPUNEWS and EPUGEN**

| From | To | Jan 1996-May 2020 | Jan1996-Dec 2007 | Jan 2008May2020 |
|---|---|---|---|---|
| UNCERT | NEWS | 0.017 | 0.000 | 0.001 |
| NEWS | UNCERT | 0.38 | 0.78 | 0.67 |
| UNCERT | GEN | 0.010 | 0.029 | 0.014 |
| GEN | UNCERT | 0.38 | 0.63 | 0.29 |

The results are unequivocal. The UNCERT series Granger causes both EPU series, and the EPU series do not Granger cause UNCERT.

The series UNCERT is based on the word "uncertainty", with a small number of words added which are closest to it, as defined by results obtained using a machine learning-based methodology.

In terms of the EPUNEWS variable, the website states that "we search for articles containing the term 'uncertainty' or 'uncertain', the terms 'economic' or 'economy' and one or more of the following terms: 'congress', 'legislation', 'white house', 'regulation', 'federal reserve', or 'deficit'".

In other words, the core word of each is the same, but the additional information used is different. The results suggest that policy makers react to changes in uncertainty rather than anticipating them.

**Appendix**

10 most similar words to uncertainty:

uncertainty, uncertainties, uncertain, unpredictability, ambiguity, certainty, confusion, turmoil, expectation, instability

R packages used for the Granger causality tests:

- *tseries* – we use the two functions *adf.test* and *kpss.test* (the Augmented Dickey-Fuller test and Kwiatkowski-Phillips-Schmidt-Shin test respectively) to check if series are stationary or contain unit roots
- *vars* – we use the function *VARselect* to compute the Akaike Information Criteria for VAR(p) processes with p from 1 through 20. We use the *VAR* function for estimating a VAR(p) process. We use the function *serial.test* to compute the multivariate Portmanteau- and Breusch-Godfrey tests for serially correlated errors in a VAR(p) process. We use the function *stability* to compute empirical fluctuation processes according to the OLS-CUSUM method
- *aod* – we use the function *wald.test* to perform the Wald tests for Granger causality

The order of integration of each series is one, both over the full sample and over each sub-period.

Number of lags selected using Akaike Information Criteria varying the number of lags from 1 through 10:

UNCERT and EPUNEWS, January 1996 – May 2020

|        | 1 | 2 | 3 | 4 | 5 | 6 | 7 | 8 | 9 | 10 |
|--------|---|---|---|---|---|---|---|---|---|----|
| AIC(n) | -1.8987518 | -1.9213758 | -1.9536281 | -1.971118 | -1.9555251 | -1.9366842 | -1.9377895 | -1.9527367 | -1.9503500 | -1.925942 |
| HQ(n)  | -1.8677617 | -1.8697256 | -1.8813179 | -1.878148 | -1.8418948 | -1.8023939 | -1.7828391 | -1.7771263 | -1.7540795 | -1.709012 |
| SC(n)  | -1.8214632 | -1.7925614 | -1.7732880 | -1.739252 | -1.6721335 | -1.6017668 | -1.5513464 | -1.5147678 | -1.4608554 | -1.384922 |
| FPE(n) | 0.1497557 | 0.1464065 | 0.1417617 | 0.139307 | 0.1415013 | 0.1441999 | 0.1440506 | 0.1419263 | 0.1422818 | 0.145818 |

UNCERT and EPUGEN, January 1996 – May 2020

|        | 1 | 2 | 3 | 4 | 5 | 6 | 7 | 8 | 9 | 10 |
|--------|---|---|---|---|---|---|---|---|---|----|
| AIC(n) | -2.92967022 | -2.95859245 | -2.99209591 | -3.00358143 | -2.99423056 | -2.98138850 | -2.97003639 | -2.96581943 | -2.95211763 | -2.93232283 |
| HQ(n)  | -2.89868014 | -2.90694232 | -2.91978572 | -2.91061119 | -2.88060027 | -2.84709816 | -2.81508599 | -2.79020898 | -2.75584713 | -2.71539228 |
| SC(n)  | -2.85238159 | -2.82977807 | -2.81175578 | -2.77171555 | -2.71083893 | -2.64647112 | -2.58359325 | -2.52785054 | -2.46262299 | -2.39130244 |
| FPE(n) | 0.05341474 | 0.05189229 | 0.05018316 | 0.04961121 | 0.05007907 | 0.05072892 | 0.05131165 | 0.05153316 | 0.05225012 | 0.05330228 |

UNCERT and EPUNEWS, January 1996 – Dec 2007

|        | 1          | 2          | 3          | 4          | 5          | 6          | 7          | 8          | 9          | 10         |
|--------|------------|------------|------------|------------|------------|------------|------------|------------|------------|------------|
| AIC(n) | -3.5095672 | -3.46204655 | -3.48680294 | -3.45765439 | -3.42796469 | -3.37760310 | -3.32302322 | -3.33224611 | -3.29960736 | -3.34310398 |
| HQ(n)  | -3.4568394 | -3.37416681 | -3.36377130 | -3.29947086 | -3.23462926 | -3.14911578 | -3.05938401 | -3.03345500 | -2.96566435 | -2.97400907 |
| SC(n)  | -3.3798132 | -3.24578985 | -3.18404355 | -3.06839233 | -2.95219995 | -2.81533567 | -2.67425312 | -2.59697333 | -2.47783189 | -2.43482583 |
| FPE(n) | 0.0299103  | 0.03136768 | 0.03060437 | 0.03151633 | 0.03247697 | 0.03417097 | 0.03611168 | 0.03581111 | 0.03703989 | 0.03551163 |

UNCERT and EPUNEWS, January 2008 – May 2020

|        | 1          | 2          | 3          | 4          | 5          | 6          | 7          | 8          | 9          | 10         |
|--------|------------|------------|------------|------------|------------|------------|------------|------------|------------|------------|
| AIC(n) | -1.2772403 | -1.2823986 | -1.2977706 | -1.3418258 | -1.3031740 | -1.2885515 | -1.2812262 | -1.2452721 | -1.2205305 | -1.1759324 |
| HQ(n)  | -1.2257658 | -1.1966078 | -1.1776635 | -1.1874024 | -1.1144342 | -1.0654954 | -1.0238538 | -0.9535833 | -0.8945255 | -0.8156110 |
| SC(n)  | -1.1505724 | -1.0712854 | -1.0022121 | -0.9618220 | -0.8387249 | -0.7396571 | -0.6478865 | -0.5274871 | -0.4183002 | -0.2892568 |
| FPE(n) | 0.2788094  | 0.2773884  | 0.2731866  | 0.2614629  | 0.2718484  | 0.2759722  | 0.2781656  | 0.2885720  | 0.2960916  | 0.3099729  |

UNCERT and EPUGEN, January 1996 – Dec 2007

|        | 1           | 2           | 3           | 4           | 5           | 6           | 7           | 8           | 9           | 10          |
|--------|-------------|-------------|-------------|-------------|-------------|-------------|-------------|-------------|-------------|-------------|
| AIC(n) | -4.45074704 | -4.40654136 | -4.43530439 | -4.39908607 | -4.37218642 | -4.32340776 | -4.26745650 | -4.24378689 | -4.22980182 | -4.33176956 |
| HQ(n)  | -4.39801919 | -4.31866162 | -4.31227276 | -4.24090254 | -4.17885099 | -4.09492044 | -4.00381728 | -3.94499577 | -3.89585881 | -3.96267466 |
| SC(n)  | -4.32099302 | -4.19028466 | -4.13254501 | -4.00982401 | -3.89642168 | -3.76114034 | -3.61868639 | -3.50851410 | -3.40802636 | -3.42349141 |
| FPE(n) | 0.01167002  | 0.01219814  | 0.01185371  | 0.01229354  | 0.01263296  | 0.01327088  | 0.01404383  | 0.01439265  | 0.01461138  | 0.01321291  |

UNCERT and EPUGEN, January 2008 – May 2020

|        | 1           | 2           | 3           | 4           | 5           | 6           | 7           | 8          | 9          | 10         |
|--------|-------------|-------------|-------------|-------------|-------------|-------------|-------------|------------|------------|------------|
| AIC(n) | -2.34533664 | -2.34119323 | -2.38227773 | -2.41026714 | -2.37937718 | -2.37986164 | -2.33887542 | -2.2883201 | -2.2701421 | -2.2434799 |
| HQ(n)  | -2.29386216 | -2.25540243 | -2.26217061 | -2.25584370 | -2.19063742 | -2.15680555 | -2.08150302 | -1.9966314 | -1.9441370 | -1.8831585 |
| SC(n)  | -2.21866870 | -2.13008000 | -2.08671920 | -2.03026332 | -1.91492807 | -1.83096723 | -1.70553572 | -1.5705351 | -1.4679118 | -1.3568043 |
| FPE(n) | 0.09581623  | 0.09621874  | 0.09235576  | 0.08982389  | 0.09266969  | 0.09266494  | 0.09659889  | 0.1016867  | 0.1036539  | 0.1065844  |

Degrees of freedom in Wald tests:

UNCERT and EPUNEWS, January 1996 – May 2020    4

UNCERT and EPUGEN, January 1996 – May 2020     4

UNCERT and EPUNEWS, January 1996 – Dec 2007    1

UNCERT and EPUNEWS, January 2008 – May 2020    4

UNCERT and EPUGEN, January 1996 – Dec 2007     1

UNCERT and EPUGEN, January 2008 – May 2020     4